# Set Theory for Verification: I. From Foundations to Functions*

*Lawrence C. Paulson*
Computer Laboratory
University of Cambridge

**Abstract**

A logic for specification and verification is derived from the axioms of Zermelo-Fraenkel set theory. The proofs are performed using the proof assistant Isabelle. Isabelle is **generic**, supporting several different logics. Isabelle has the flexibility to adapt to variants of set theory. Its higher-order syntax supports the definition of new binding operators. Unknowns in subgoals can be instantiated incrementally. The paper describes the derivation of rules for descriptions, relations and functions, and discusses interactive proofs of Cantor's Theorem, the Composition of Homomorphisms challenge [9], and Ramsey's Theorem [5]. A generic proof assistant can stand up against provers dedicated to particular logics.

*Research funded by the SERC (grants GR/G53279, GR/H40570) and by the ESPRIT Basic Research Actions 3245 'Logical Frameworks' and 6453 'Types'. Isabelle has enjoyed long-standing support from the British SERC, dating from the Alvey Programme (grant GR/E0355.7).

# Contents

# 1 Introduction

Type theory has been highly successful in specification and verification. Higher-order logic (HOL) is the most popular form of type theory [12]; consider for example Graham's correctness proofs about an SECD microprocessor [20] using Gordon's HOL system [18, 19]. More recent type theories, based on the propositions-as-types principle, are being applied to the problem of deriving programs from specifications. Thompson [52] presents examples using Martin-Löf Type Theory [33], while Leclerc and Paulin's work on streams [27] is a typical application of the Calculus of Constructions.

Type theory contains a kind of set theory. A term $p$ of type $\tau \Rightarrow \mathit{bool}$ represents a set, and the formula $p(a)$ represents $a \in p$. Intersections, unions and so on are easily defined. But this is a typed set theory. Every element of $p$ must have type $\tau$. Lamport [26] argues that types are harmful in specifications; I regard his view as unproven but deserving investigation. Set theory — the traditional, untyped version — has seldom been used for verification. An implementation of set theory could support untyped specification languages such as TLA [25].

Set theory is harder to automate than type theory. It is low-level, with strange definitions like $\langle a, b\rangle \equiv \{\{a\}, \{a, b\}\}$ and $3 \equiv \{0, 1, 2\}$; and since it has no information hiding, it admits strange theorems like $\{a\} \in \langle a, b\rangle$ and $2 \in 3$. The search space is larger in set theory than in type theory: set theory proofs need extra steps to perform type checking, and the lack of type constraints admits more terms as well-formed. Virtually all type theories satisfy strong normalization: every reduction sequence terminates in a normal form. The corresponding notion of reduction in set theory, which reduces $t \in \{x . \psi[x]\}$ to $\psi[t]$, admits expressions with no normal form [3]. For example, $\{x . x \in x\} \in \{x . x \in x\}$ reduces to itself [10, page 295]. Type theory provers rely upon strong normalization; a set theory prover must do without this property.

Many existing set theory provers (surveyed in §11) employ domain-dependent knowledge and aim to model the reasoning methods of mathematicians. The present work aims to provide a productive environment for verification. While Part I of this paper treats familiar mathematical examples (Cantor's Theorem and Ramsey's Theorem), most of the effort in Part II [40] concerns lists and other computational theories.

The implementation is based on Isabelle, a generic proof assistant [37]. No other set theory prover simultaneously supports higher-order logic (HOL), Constructive Type Theory (CTT), the sequent calculus (LK), various modal logics, and the Logic for Computable Functions (LCF). Set theory, starting with the Zermelo-Fraenkel axioms (ZF), is built on Isabelle's implementation of classical first-order logic (FOL). The latter is built on Isabelle's intuitionistic first-order logic (IFOL). Figure 1 displays some of Isabelle's predefined object-logics; users may define new ones.

A generic theorem prover offers flexibility as logical formalisms evolve. Set theory has remained stable for decades; now variants are being proposed, such as non-well-founded set theory [1] and intuitionistic set theory (IZF) [28]. Modifying a specialized set theory prover in order to support IZF would be difficult. With

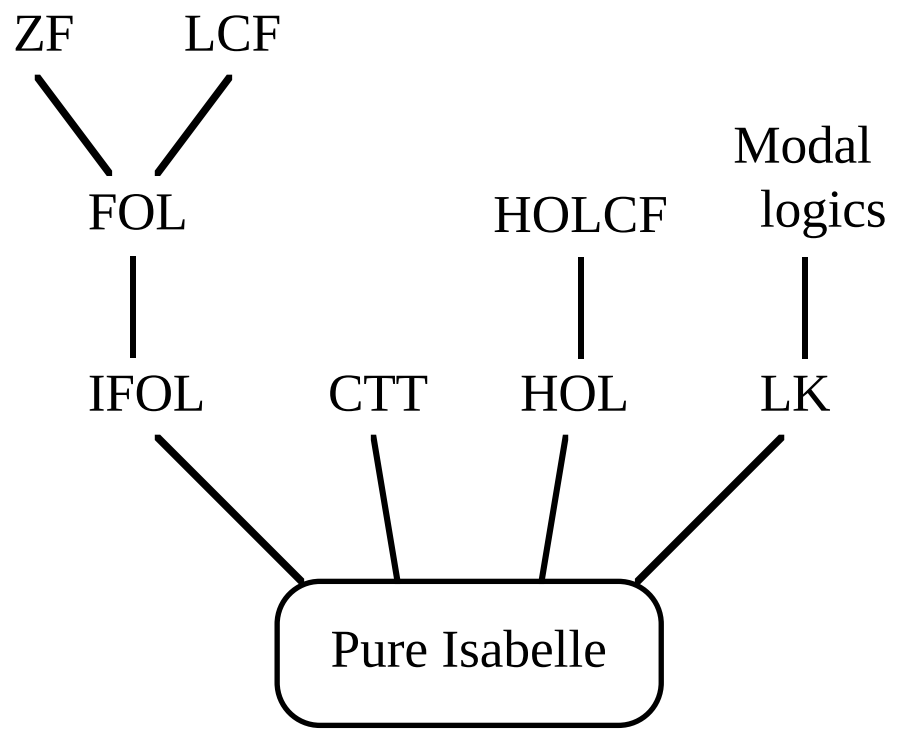

Figure 1: Isabelle's object logics

Isabelle, we simply copy the existing definition of ZF, modify the axioms as proposed, and base the new theory upon IFOL instead of FOL. Many IZF proofs could be obtained by modifying the corresponding ZF ones.

Isabelle's generic facilities help in other ways. Its higher-order syntax handles variable-binding operators such as $\{x \in A . \psi[x]\}$ and $\iota x . \psi[x]$. Its parser and pretty printer helps give these operators a readable concrete syntax. Isabelle handles axiom and theorem schemes, which abound in set theory. Isabelle's version of set theory includes derived theories of relations, functions and type constructions, which can be regarded as logics in their own right.

My work incorporates some of Noël's [32]. Noël proved a large collection of examples by proof checking, demonstrating that Isabelle could cope with set theory. My contribution is to greatly improve Isabelle's level of automation, and to derive a theory that can serve as a platform for verifications.

This paper can be seen as an extended introduction to Isabelle. It describes how a logic is defined, how theorems are proved, and how theories are developed. Part I of the paper derives a computational logic inspired by Martin-Löf Type Theory [33]. Part II continues the development to handle recursion [40].

Part I proceeds as follows. The next two sections introduce Isabelle and axiomatic set theory. Further sections sketch the Isabelle development of basic concepts such as relations and functions. Next come interactive proofs of three small examples: ordered pairing, Cantor's Theorem, and the Composition of Homomorphisms challenge [9]. Ramsey's Theorem, a more realistic example, permits a comparison between Isabelle and other theorem provers [5]. The remaining sections discuss related work and draw conclusions.

# 2 Isabelle

Here is a brief survey of Isabelle. For more information, please see the documentation [37, 38, 39]. Elsewhere [35] I discuss how to formalize object-logics in the meta-logic, and how to prove that the formalization is correct.

## 2.1 The meta-logic

Isabelle works directly with schematic inference rules of the form

$$[\![\phi_1; \ldots; \phi_n]\!] \Longrightarrow \phi.$$

Rules are combined by a generalization of Horn clause resolution. Theorems are proved not by refutation, but in the affirmative. Joining rules by resolution constructs a proof tree, whose root is the conclusion.

Such rules are theorems of Isabelle's meta-logic, which is a fragment of intuitionistic higher-order logic. The $\phi_i$ are formulae of the meta-logic, and are called **propositions**. Atomic propositions assert judgements of an object-logic, such as '$P$ is true', '$A$ is a well-formed type', or '$a$ has type $A$'. These typically involve user-declared predicates. Propositions are combined by the meta-level connectives $\Longrightarrow$, $\bigwedge$ and $\equiv$.

The connective $\Longrightarrow$ is meta-implication. It expresses entailment in rules, including assumption discharge for natural deduction. The notation $[\![\phi_1; \ldots; \phi_n]\!] \Longrightarrow \phi$ abbreviates

$$\phi_1 \Longrightarrow (\cdots \Longrightarrow (\phi_n \Longrightarrow \phi) \cdots).$$

The connective $\bigwedge$ is the meta-level universal quantifier. It expresses generality in rules. Locally $\bigwedge$-bound variables represent eigenvariables, formalizing the '$x$ not free in ...' provisos typical of quantifier rules [35]. Outermost $\bigwedge$-bound variables are schematic, although such quantifiers are usually dropped.

The connective $\equiv$ is meta-equality. It expresses definitions.

The meta-logic includes a rule for $\equiv$, which expands definitions explicitly. Although there are separate rules for $\Longrightarrow$ and $\bigwedge$, these connectives are also 'hard-wired' into Isabelle's resolution rule.

## 2.2 Variable binding

Expressions in the meta-logic are typed $\lambda$-terms, and $\lambda$-abstraction handles an object-logic's quantifiers and other variable-binding operators. Let us distinguish between substitution and application. Writing $P[x]$ emphasizes the free occurrences of the variable $x$ in $P$; in this context, $P[t]$ stands for the result of substituting $t$ for $x$ in $P$. On the other hand, $P(t)$ simply stands for the application of $P$ to $t$ in Isabelle's $\lambda$-calculus. For $P(t)$ to be well-typed, $P$ must have type $\sigma \Rightarrow \tau$ and $t$ must have type $\sigma$. In short, $P[t]$ stands for **substitution** while $P(t)$ stands for **application**. Because Isabelle reduces $(\lambda x \,.\, P[x])(t)$ to $P[t]$, Isabelle theory definitions can express substitution by application.

In Isabelle's set theory, where $i$ is the type of sets and $o$ is the type of formulae, the quantifiers could be declared[1] to have the type $(i \Rightarrow o) \Rightarrow o$:

$$\forall, \exists \;\; :: \;\; (i \Rightarrow o) \Rightarrow o$$

[1]They actually have the polymorphic type $(\alpha \Rightarrow o) \Rightarrow o$. Isabelle's *type classes* regulate polymorphism [37].

If $P[x]$ is a formula (represented by a term of type $o$) and $x$ is a variable of type $i$, then $\lambda x\,.\,P[x]$ has type $i \Rightarrow o$. Thus $\forall(\lambda x\,.\,P[x])$ has type $o$, and represents the formula $\forall x.P[x]$. The description operator $\iota$ has type $(i \Rightarrow o) \Rightarrow i$; note that $\iota x.P[x]$ is a term rather than a formula. Function abstraction in set theory is represented by the constant `Lambda` of type $[i, i \Rightarrow i] \Rightarrow i$, where $\texttt{Lambda}(A, \lambda x\,.\,b[x])$ denotes the function that returns $b[a]$ for all $a \in A$. This paper writes $\lambda_{x \in A}\,.\,b[x]$ instead of $\texttt{Lambda}(A, \lambda x\,.\,b[x])$ below, using the symbol $\lambda$ with two distinct meanings.

Isabelle's $\lambda$-calculus allows schematic definitions of variable-binding operators. Consider bounded quantification. Let `Ball` be a constant of type $[i, i \Rightarrow o] \Rightarrow o$. We may define

$$\texttt{Ball}(A, P) \;\equiv\; \forall x\,.\,x \in A \to P(x)$$

where $P$ is a *variable* of type $i \Rightarrow o$, standing for any unary property of sets. This paper writes $\forall_{x \in A}\,.\,P(x)$ instead of $\texttt{Ball}(A, P)$ below.

## 2.3 Basic proof methods

Isabelle uses **higher-order unification** [22], rather than ordinary unification, to find unifiers in the typed $\lambda$-calculus. Higher-order unification is undecidable in the general case but works well on the simpler cases that arise in practice. In particular, it handles quantifier reasoning [35]. Unknowns may be shared among subgoals and may be instantiated incrementally.

For backward proof, a rule of the form $[\![\phi_1; \ldots; \phi_n]\!] \Longrightarrow \phi$ can represent a **proof state**; the ultimate goal is $\phi$ and the subgoals still unsolved are $\phi_1, \ldots, \phi_n$. An **initial** proof state has the form $\phi \Longrightarrow \phi$, with one subgoal, and a **final** proof state has the form $\phi$, with no subgoals. The final proof state *is itself* the desired theorem.

**Tactics** are functions that transform proof states. A backward proof proceeds by applying tactics in succession to the initial state, reaching a final state. The tactic `resolve_tac` performs Isabelle's version of Horn clause resolution; it attempts to unify the conclusion of some inference rule with a subgoal, replacing it by the rule's instantiated premises. This is proof checking.

With derived rules, even proof checking can be effective. For example, three derived rules express that $A \cap B$ is the greatest lower bound of $A$ and $B$:

$$A \cap B \subseteq A \qquad A \cap B \subseteq B \qquad \frac{C \subseteq A \quad C \subseteq B}{C \subseteq A \cap B}$$

Another derived rule states that the $\wp$ operator, where $\wp(A)$ is the set of all subsets of $A$, is monotonic:

$$\frac{A \subseteq B}{\wp(A) \subseteq \wp(B)}$$

These derived rules yield a simple proof of $\wp(A \cap B) \subseteq \wp(A) \cap \wp(B)$; try it. In Isabelle, derived rules are indistinguishable in use from primitive ones.[2]

[2]The Isabelle Logics Manual [39, §3.9] shows an interactive proof of $\wp(A \cap B) = \wp(A) \cap \wp(B)$. This session contains the proof mentioned above; the opposite inclusion is equally elegant.

## 2.4 Automated reasoning

Each tactic maps a proof state to a **lazy list** of possible next states. Backtracking is therefore possible and tactics can implement search strategies such as depth-first, best-first and iterative deepening. **Tacticals** are operators for combining tactics. They typically express control structures, ranging from basic sequencing to search strategies. Isabelle provides several powerful, generic tools:

- The **classical reasoner** applies naïve heuristics to prove theorems in the style of the sequent calculus. Despite its naïveté, it can prove nearly all of Pelletier's graded problems short of Schubert's Steamroller [41]. As an interactive tool it is invaluable. It is not restricted to first-order logic, but exploits any natural deduction rules. It can prove several key lemmas for Ramsey's Theorem [5].

- The **simplifier** applies rewrite rules to a goal, then attempts to prove the rewritten goal using a user-supplied tactic. A conditional rewrite rule is applied only if recursive simplification proves the instantiated condition. Contextual information is also used, rewriting $x = t \rightarrow \psi[x]$ to $x = t \rightarrow \psi[t]$. Rewriting works not just for equality, but for any reflexive/transitive relation enjoying congruence laws. Used with the classical reasoner, it can prove Boyer et al.'s challenge problem, that the composition of homomorphisms is a homomorphism [9].

Isabelle does not find proofs automatically. Proofs require a skilled user, who must decide which lemmas to prove and which tools to apply. If no tool is appropriate, proof checking is always available.

# 3 Set theory

Axiomatic set theory was developed in response to paradoxes such as Russell's. Sets could not be arbitrary collections of the form $\{x \,.\, \phi[x]\}$, pulled out of a hat. They had to be constructed, starting from a few given sets. Operations for constructing new sets included union, powerset and replacement.

**Replacement** is the most powerful set constructor. Its parameters are a set $A$ and a binary predicate $\phi(x, y)$ such that

$$\forall_{x \in A} \,.\, \forall y\, z \,.\, \phi(x, y) \wedge \phi(x, z) \rightarrow y = z.$$

(We say that $\phi$ is **single-valued** for $x$ in $A$.) Replacement yields the image of $A$ under the predicate $\phi$: there exists a set $B$, depending upon $A$ and $\phi$, such that $b \in B \leftrightarrow (\exists_{x \in A} \,.\, \phi(x, b))$. The Isabelle formalization declares a constant $\mathcal{R}$ of type $[i, [i, i] \Rightarrow o] \Rightarrow i$; the replacement axiom concludes

$$b \in \mathcal{R}(A, \phi) \leftrightarrow (\exists_{x \in A} \,.\, \phi(x, b)).$$

The predicate $\phi$ typically has the form $\lambda x\, y \,.\, P$; thus $\mathcal{R}$ is a binding operator.

Replacement entails the principle of **Separation**. Let $A$ be a set and $\psi[x]$ a unary predicate. Separation yields a set, written $\{x \in A \,.\, \psi[x]\}$, consisting of those elements of $A$ that satisfy $\psi$:

$$a \in \{x \in A \,.\, \psi[x]\} \leftrightarrow a \in A \wedge \psi[a]$$

The **class** $\{x \,.\, \psi[x]\}$ is an unrestricted collection of sets. Every set $B$ is a class, namely $\{x \,.\, x \in B\}$. Many classes are too big to be sets, such as the **universal class**, $V \equiv \{x \,.\, x = x\}$. If $V$ were a set then we could obtain Russell's Paradox via Separation: define the set $R \equiv \{x \in V \,.\, x \notin x\}$, then $R \in R \leftrightarrow R \notin R$. We could define $R$ as a class, namely $R \equiv \{x \,.\, x \notin x\}$, but this yields no paradox because a proper class cannot be a member of another class: $R \in R$ is false.

## 3.1 Which axiom system?

The two main axiom systems for set theory, Zermelo-Fraenkel (ZF) and Bernays-Gödel (BG), differ in their treatment of classes. In ZF, variables range over sets; classes do not exist at all, but we may regard unary predicates as classes if we like. In BG, variables range over classes, and $A \in V$ states that the class $A$ is actually a set. Although ZF and BG are similar in strength, mathematicians generally consider ZF to be the standard system for set theory [14, page iii] [50, page 327]. BG frequently requires showing that certain classes are sets; $a \in \{a\}$ holds only if $a \in V$.

One obstacle to automating ZF is that Replacement is an infinite axiom scheme. There is no finite axiom system for ZF, if ZF is consistent [24, page 138]. Here is a proof sketch. The Reflection Theorem implies that every finite subset of the ZF axioms has a model in ZF. Assume, for contradiction, that ZF has a finite axiom system. By the Compactness Theorem, ZF has a finite axiom system $Z_0$ that is a subset of the ZF axioms. By the Reflection Theorem, $Z_0$ has a model in ZF. Thus ZF proves the consistency of $Z_0$. Now $Z_0$ implies the whole of ZF, so $Z_0$ proves the consistency of itself; this contradicts Gödel's Incompleteness Theorem.

Boyer et al. [9] advocate BG because it is a finite system of axioms; Quaife [44] has proved hundreds of theorems in BG, using the resolution prover Otter. But the existence of a finite axiom system is perhaps irrelevant. Saaltink's set theory proofs [47] employ ZF, adopting a simple device for invoking instances of axiom schemes; a resolution prover could do the same. Quaife notes that even BG requires schematic reasoning in normal use. The key issue, then, is not finiteness of the axiom system, but schematic reasoning.

## 3.2 Schematic reasoning in set theory

Isabelle supports schematic reasoning through its higher-order meta-logic. In the axiom of Replacement, the binary predicate $\phi$ is a variable of type $[i, i] \Rightarrow o$, which is the type of functions that map two sets to a truth value. Separation can be defined in its schematic form, where the unary predicate $\psi$ is a variable of type $i \Rightarrow o$. Rules about Separation can be proved schematically.

| *syntax* | *description* |
|---|---|
| `{`$a_1$`, ...,` $a_n$`}` | finite set |
| `<`$a$`,` $b$`>` | ordered pair |
| $\{x\!:\!A\,.\,P[x]\}$ | Separation |
| $\{y\,.\,x\!:\!A,\ Q[x,y]\}$ | Replacement |
| $\{b[x]\,.\,x\!:\!A\}$ | functional Replacement |
| `INT` $x\!:\!A\,.\,B[x]$ | $\bigcap_{x\in A}.B[x]$, general intersection |
| `UN` $x\!:\!A\,.\,B[x]$ | $\bigcup_{x\in A}.B[x]$, general union |
| $A$ `Int` $B$ | $A\cap B$, intersection |
| $A$ `Un` $B$ | $A\cup B$, union |
| $A$ `->` $B$ | $A\to B$, function space |
| $A$ `*` $B$ | $A\times B$, Cartesian product |
| `PROD` $x\!:\!A\,.\,B[x]$ | $\Pi_{x\in A}\,.\,B[x]$, general product |
| `SUM` $x\!:\!A\,.\,B[x]$ | $\Sigma_{x\in A}\,.\,B[x]$, general sum |
| `THE` $x\,.\,P[x]$ | $\iota x\,.\,P[x]$, definite description |
| `lam` $x\!:\!A\,.\,b[x]$ | $\lambda_{x\in A}\,.\,b[x]$, abstraction |
| $a$ `:` $A$ | $a\in A$, membership |
| $A$ `<=` $B$ | $A\subseteq B$, subset relation |
| `ALL` $x\!:\!A\,.\,P[x]$ | $\forall_{x\in A}\,.\,P[x]$, bounded quantifier |
| `EX` $x\!:\!A\,.\,P[x]$ | $\exists_{x\in A}\,.\,P[x]$, bounded quantifier |

Figure 2: ASCII notation for ZF

Schemes can, however, make the search space too big. The goal

$$t \in {?A},$$

could be refined, instantiating the unknown $?A$ (a 'logical variable'), to the subgoal

$$t \in \{x \in {?B}\,.\,{?\psi}(x)\}.$$

This can be refined, by the principle of Separation, to the two subgoals

$$t \in {?B} \qquad \text{and} \qquad {?\psi}(t).$$

The subgoal $t \in {?B}$ can be refined exactly like $t \in {?A}$, making the search diverge; the subgoal $?\psi(t)$ is totally unconstrained, since it consists of a formula unknown. Had we proved $t \in {?A}$ by instantiating $?A$ to $\{t\}$, we might have invalidated other goals involving $?A$. Such a situation arises in the proof of Cantor's Theorem (see §8.2). Automatic tools seldom cope; the user can help by explicitly instantiating unknowns such as $?A$, or by artificially restricting the search.

# 4 The Zermelo-Fraenkel axioms in Isabelle

The ZF axioms from Suppes [51, page 238] are expressed using Isabelle's formulation of classical first-order logic. For clarity, the exposition uses standard mathematical

```
Ball_def     "Ball(A,P) == ALL x. x:A --> P(x)"
Bex_def      "Bex(A,P) == EX x. x:A & P(x)"
subset_def   "A <= B == ALL x:A. x:B"
extension    "A = B <-> A <= B & B <= A"
union_iff    "A : Union(C) <-> (EX B:C. A:B)"
power_set    "A : Pow(B) <-> A <= B"
foundation   "A=0 | (EX x:A. ALL y:x. ~ y:A)"

replacement "(ALL x:A. ALL y z. P(x,y) & P(x,z) --> y=z) ==> \
\            b : PrimReplace(A,P) <-> (EX x:A. P(x,b))"
```

Figure 3: The ZF axioms from the Isabelle theory file

notation rather than Isabelle's ASCII substitutes (Figure 2). We begin by defining the bounded quantifiers:

$$\begin{aligned} \forall_{x\in A} . \psi(x) &\equiv \forall x . x \in A \to \psi(x) \\ \exists_{x\in A} . \psi(x) &\equiv \exists x . x \in A \wedge \psi(x) \end{aligned}$$

Taking membership ($\in$) as a primitive binary relation, we define the subset relation:

$$A \subseteq B \equiv \forall_{x\in A} . x \in B$$

The following axioms are standard:

$$\begin{aligned} A = B &\leftrightarrow A \subseteq B \wedge B \subseteq A && \text{(Extensionality)} \\ A \in \bigcup(C) &\leftrightarrow (\exists_{B\in C} . A \in B) && \text{(Union)} \\ A \in \wp(B) &\leftrightarrow A \subseteq B && \text{(Powerset)} \\ A = \emptyset &\vee (\exists_{x\in A} . \forall_{y\in x} . y \notin A) && \text{(Foundation)} \end{aligned}$$

Replacement is expressed by a rule whose premise asserts that $\phi$ is single-valued:

$$\frac{\forall_{x\in A} . \forall y\, z . \phi(x,y) \wedge \phi(x,z) \to y = z}{b \in \mathcal{R}(A,\phi) \leftrightarrow (\exists_{x\in A} . \phi(x,b))} \qquad \text{(Replacement)}$$

These are all the axioms apart from Infinity, which is not discussed in this paper, and Choice, which I have not used at all. Figure 3 presents the axioms as they appear in the actual Isabelle theory definition. The constant `PrimReplace` is just $\mathcal{R}$, while `Ball` and `Bex` are used in the internal representation of the bounded quantifiers. The file contains many other definitions. Isabelle accepts infix and mixfix declarations for defining new notation, hiding the internal representation. The parser and pretty printer need approximately 100 lines of ML to specify the translation of set theory's binding operators and other notation shown in Figure 2.

Is this Isabelle formalization faithful to ZF? I am confident that it is, although I have not proved this formally. Let us consider some of the issues.

*Notation.* According to Devlin [14, page 36], the formal language of ZF consists of first-order formulae built up from $x = y$ and $x \in y$; the only terms are variables. The Isabelle theory defines a rich language; its Replacement notation $\mathcal{R}(A, \phi)$ has no counterpart in most texts. Extending the language risks making the theory stronger than ZF, since it creates additional instances of axiom schemes.

But the additional notation can be eliminated. If we can replace each formula of the form $x \in C$ by an equivalent formula not mentioning $C$, then we can eliminate all mention of $C$. The Union and Powerset Axioms describe how to eliminate the symbols $\bigcup$ and $\wp$. My Replacement Axiom, following tradition, takes single-valued($\phi$) as a condition; it says nothing about $\mathcal{R}(A, \phi)$ when $\phi$ is not single-valued. We can strengthen the axiom to force $\mathcal{R}(A, \phi) = \emptyset$ in that case:

$$b \in \mathcal{R}(A, \phi) \leftrightarrow (\exists_{x \in A} \,.\, \phi(x, b)) \land \text{single-valued}(\phi) \qquad \text{(Replacement')}$$

This stronger Axiom shows that the notation $\mathcal{R}(A, \phi)$ can be eliminated from all formulae and is therefore harmless.

*Soundness of the meta-logic.* I have previously proved the correctness of Isabelle's formulation of intuitionistic first-order logic [35], exhibiting a strong correspondence between meta-level and object-level proofs. By a standard result of proof theory, any proof in Isabelle's meta-logic can be put into normal form [43]. Given a normal meta-proof, we can read off the corresponding object-proof. The correspondence must be shown for each inference rule of the object-logic, but the argument is similar in each case. The approach works for all rules of the same general form, whose premises may discharge assumptions and have eigenvariables. It does not work for linear, relevance and modal logics, whose rules violate the structural framework of natural deduction (for instance, by stipulating that proofs must use each assumption).

The ZF formalization extends intuitionistic FOL with the double-negation rule (to obtain classical logic) and the ZF axioms. Let us continue to use the stronger form of Replacement. The previously demonstrated correspondence [35] easily accommodates these extensions; the double-negation rule is a simple example of assumption discharge, while the ZF axioms and axiom schemes are mere formulae. Semantically this may seem strange — there is a world of difference between intuitionistic FOL and set theory — but proofs in both systems share the same syntactic structure. Isabelle's approach is unconcerned with models and can even handle inconsistent object-logics. We could formalize naive set theory in Isabelle's meta-logic and expect all proofs, including Russell's Paradox, to be faithfully represented.

*Pragmatics.* Obviously Isabelle itself contains bugs, since it is a large unverified program. The distinction between theory and implementation is not entirely clear; for instance, the Isabelle theory of FOL includes some constants and axioms whose sole purpose is to speed rewriting. There are many such details that look harmless and are too tiresome to consider explicitly. Although I have no comprehensive correctness proof for the Isabelle version of ZF, its theoretical basis appears to be firm.

# 5 Natural deduction rules for set theory

The theory above is largely in the form of logical equivalences; perhaps we could develop a transformational calculus. I prefer to derive natural deduction rules. In natural deduction, each rule introduces or eliminates some constant. Repeatedly applying such rules breaks down a formula to atomic formulae; this can be automated, yielding a proof procedure resembling those for semantic tableaux. Such a proof procedure lies at the heart of Isabelle's classical reasoner.

From the definition $A \subseteq B \equiv \forall_{x \in A} . x \in B$ we obtain introduction and elimination rules for $\subseteq$:

$$\frac{\begin{matrix}[x \in A]_x \\ \vdots \\ x \in B\end{matrix}}{A \subseteq B}\ (\subseteq I) \qquad \frac{A \subseteq B \quad c \in A}{c \in B}\ (\subseteq E)$$

Rule $(\subseteq I)$ discharges the assumption $x \in A$; it holds provided $x$ is not free in the conclusion or other assumptions. Here and below, premises indicate such provisos by subscripting the affected variable.

From the Union axiom, we obtain introduction and elimination rules for $\bigcup$:

$$\frac{B \in C \quad A \in B}{A \in \bigcup(C)}\ (\bigcup I) \qquad \frac{A \in \bigcup(C) \qquad \begin{matrix}[A \in X \quad X \in C]_X \\ \vdots \\ \theta\end{matrix}}{\theta}\ (\bigcup E)$$

Rule $(\bigcup E)$ discharges two assumptions, and has another 'not free' proviso on $X$.

Natural deduction rules are more compact than sequent calculus rules because they leave the context implicit: each rule mentions only the assumptions it discharges. They are certainly preferable to expanding the definitions of the constants. The style constrains the form of each rule and provides a naming convention. Binary intersection has a rather technical definition (§7.1). Its natural deduction rules are straightforward:

$$\frac{c \in A \quad c \in B}{c \in A \cap B}\ (\cap I) \qquad \frac{c \in A \cap B}{c \in A}\ (\cap E1) \qquad \frac{c \in A \cap B}{c \in B}\ (\cap E2)$$

For instance, $A \cap B = B \cap A$ has a simple proof using these rules. Schmidt [49] also advocates natural deduction in set theory.

Isabelle uses natural deduction rules for reasoning backwards from the goal and for reasoning forwards from the assumptions. The rules $(\subseteq I)$ and $(\subseteq E)$ are formalized as the following theorems of Isabelle's meta-logic. In $(\subseteq I)$, observe the use of $\bigwedge$ to introduce an eigenvariable, and the use of $\Longrightarrow$ to discharge an assumption [35].

$$(\bigwedge x \,.\, x \in A \Longrightarrow x \in B) \Longrightarrow A \subseteq B \qquad (\subseteq I)$$

$$[\![A \subseteq B;\ c \in A]\!] \Longrightarrow c \in B \qquad (\subseteq E)$$

In Isabelle's ASCII syntax, they look like this:

```
(!!x.x:A ==> x:B) ==> A <= B
[| A <= B;  c:A |] ==> c:B
```

The Isabelle version of ZF derives natural deduction rules for all the primitives of set theory, including all the rules shown in this paper.

# 6 From Replacement to Separation

Replacement is powerful but cumbersome. This section discusses three simpler forms of Replacement. The first is as powerful as the original, but facilitates reasoning about the single-valued property. The other two are special cases of Replacement.

## 6.1 A simpler single-valued property

The introduction and elimination rules for $b \in \mathcal{R}(A, \phi)$ both require an additional premise stating that $\phi$ is single-valued. Defining a new form of Replacement reduces this proof burden. If $\phi(x, y)$ is a binary predicate, then let

$$\phi'(x, y) \quad\equiv\quad (\exists! z \,.\, \phi(x, z)) \wedge \phi(x, y).$$

Since $\exists! z \,.\, \phi(x, z)$ means there exists a *unique* $z$ such that $\phi(x, z)$, the definition ensures that $\phi'(x, y)$ is single-valued. Moreover, if $\phi(x, y)$ is already single-valued then the two predicates are equivalent. Isabelle expresses $\phi'$ in terms of $\phi$ using meta-level $\lambda$-abstraction; we may define the new form of Replacement (with a nice notation) by

$$\{y \,.\, x \in A, \phi(x, y)\} \quad\equiv\quad \mathcal{R}(A,\ \lambda x\, y \,.\, (\exists! z \,.\, \phi(x, z)) \wedge \phi(x, y)).$$

We easily obtain the equivalence

$$b \in \{y \,.\, x \in A, \phi(x, y)\} \leftrightarrow (\exists_{x \in A} \,.\, \phi(x, b) \wedge (\forall y \,.\, \phi(x, y) \rightarrow y = b)).$$

This equivalence is unconditional. It never asks whether $\phi(x, y)$ is single-valued for all $x$ in $A$, only for some value of $x$ such that $\phi(x, b)$.

Using the new definition, we derive natural deduction rules. The introduction rule includes a simplified premise about the single-valued property. The elimination rule requires no such premise; on the contrary, it discharges an assumption involving this property. (The assumption, omitted below for clarity, is $\forall y \,.\, \phi(x, y) \rightarrow y = b$.)

$$\frac{a \in A \quad \phi(a, b) \quad \begin{array}{c}[\phi(a, y)]_y \\ \vdots \\ y = b\end{array}}{b \in \{y \,.\, x \in A, \phi(x, y)\}}\ (\mathcal{R}I) \qquad \frac{b \in \{y \,.\, x \in A, \phi(x, y)\} \quad \begin{array}{c}[x \in A \quad \phi(x, b)]_x \\ \vdots \\ \theta\end{array}}{\theta}\ (\mathcal{R}E)$$

## 6.2 Functional Replacement

Suppose that $f$ is a unary operator on sets — not a set-theoretic function, which is a set of pairs, but a meta-level function such as $\wp$ or $\bigcup$. Since the predicate $\phi(x, y) \equiv (y = f(x))$ is obviously single-valued, define

$$\{f(x) \,.\, x \in A\} \quad\equiv\quad \{y \,.\, x \in A, y = f(x)\}.$$

This form of Replacement illustrates why single-valued predicates are sometimes called **class functions**. Isabelle can express meta-level functions by abstraction in its typed $\lambda$-calculus.

Functional replacement, with the basic $\bigcup$ operator, expresses a more familiar form of union:

$$\bigcup_{x\in A} B(x) \quad\equiv\quad \bigcup(\{B(x)\,.\,x\in A\})$$

The corresponding natural deduction rules are

$$\frac{a\in A \quad b\in B(a)}{b\in(\bigcup_{x\in A}.B(x))}\,(\bigcup\mathcal{R}I) \qquad \frac{b\in(\bigcup_{x\in A}.B(x)) \quad \begin{array}{c}[x\in A \quad b\in B(x)]_x\\ \vdots\\ \theta\end{array}}{\theta}\,(\bigcup\mathcal{R}E)$$

## 6.3 Separation

Given a set $A$ and a unary predicate $\psi$, Separation yields a set consisting of those elements of $A$ that satisfy $\psi$. Separation is easily defined in terms of Replacement:

$$\{x\in A\,.\,\psi(x)\} \quad\equiv\quad \{y\,.\,x\in A, x=y\wedge\psi(x)\}$$

The natural deduction rules have simple derivations:

$$\frac{a\in A \quad \psi(a)}{a\in\{x\in A\,.\,\psi(x)\}} \qquad \frac{a\in\{x\in A\,.\,\psi(x)\}}{a\in A} \qquad \frac{a\in\{x\in A\,.\,\psi(x)\}}{\psi(a)}$$

Using Separation, we can define general intersection:

$$\bigcap(C) \quad\equiv\quad \{x\in\bigcup(C)\,.\,\forall_{Y\in C}\,.\,x\in Y\}$$

The empty intersection, $\bigcap(\emptyset)$, causes difficulties. It would like to contain everything, but there is no universal set; $\bigcap(\emptyset)$ should be undefined. But Isabelle's set theory does not formalize the notion of definedness; all terms are defined. Because $\bigcup(\emptyset)=\emptyset$, we obtain the perverse (but harmless) result $\bigcap(\emptyset)=\emptyset$.

## 6.4 The Isabelle definitions

It may be instructive to see how these concepts are defined in the Isabelle theory file for ZF. First, the file declares constants for the three set formers:[3]

```
Replace :: "[i, [i,i]=>o] => i"
RepFun  :: "[i, i=>i] => i"
Collect :: "[i, i=>o] => i"
```

[3] `Collect` has turned out to be an unfortunate name for the Separation constant — there is a variant of Replacement known as Collection.

Later, the file defines the constants as described above (note that % is ASCII for the meta-level $\lambda$ symbol):

```
Replace_def "Replace(A,P) ==
                 PrimReplace(A, %x y. (EX!z.P(x,z)) & P(x,y))"
RepFun_def  "RepFun(A,f)  == {y . x:A, y=f(x)}"
Collect_def "Collect(A,P) == {y . x:A, x=y & P(x)}"
```

These constants suffice to express the corresponding sets, but the notation leaves much to be desired. The file therefore goes on to define a readable syntax, with translations between $\{x{:}A.\ \psi[x]\}$ and `Collect(`$A$`, %`$x.\psi[x]$`)` for example.

# 7 Deriving a theory of functions

The next developments are tightly linked. We define unordered pairs, then binary unions and intersections, and obtain finite sets of arbitrary size. Then we can define descriptions and ordered pairs. Finally, we can define Cartesian products, binary relations and functions. The resulting theory includes a sort of $\lambda$-calculus with $\Pi$ and $\Sigma$ types.

## 7.1 Finite sets and the boolean operators

Unordered pairing is frequently taken as primitive, but it can be defined in terms of Replacement [51, page 237]. Observe that $\wp(\wp(\emptyset))$ contains two distinct elements, $\emptyset$ and $\wp(\emptyset)$.

$$\mathtt{Upair}(a,b) \equiv \{y\,.\,x\in\wp(\wp(\emptyset)),\ (x=\emptyset\wedge y=a)\vee(x=\wp(\emptyset)\wedge y=b)\}$$

Tedious but elementary reasoning yields the key property:

$$c\in\mathtt{Upair}(a,b)\leftrightarrow(c=a\vee c=b).$$

Now we can define binary union, intersection and (while we are at it) set difference:

$$\begin{aligned} A\cup B &\equiv \bigcup(\mathtt{Upair}(A,B)) \\ A\cap B &\equiv \bigcap(\mathtt{Upair}(A,B)) \\ A-B &\equiv \{x\in A\,.\,x\notin B\} \end{aligned}$$

Finite sets are traditionally obtained as binary unions of unordered pairs. Isabelle's treatment is inspired by Lisp. Define

$$\mathtt{cons}(a,B) \equiv \mathtt{Upair}(a,a)\cup B.$$

Thus $\mathtt{cons}(a,B)$ augments $B$ with the element $a$; we obtain

$$c\in\mathtt{cons}(a,B)\leftrightarrow(c=a\vee c\in B).$$

In Isabelle, the notation $\{a_1,\ldots,a_n\}$ expands to $\mathtt{cons}(a_1,\ldots,\mathtt{cons}(a_n,\emptyset)\ldots)$.

## 7.2 Descriptions

Compared with Suppes [51], Isabelle's axioms take one liberty. They do not merely assert the existence of powersets, unions and replacements, but give them names: $\wp(A)$, $\bigcup(A)$ and $\mathcal{R}(A, \phi)$. There is nothing wrong with assigning notation to objects, provided they are unique, and Suppes does so informally.

By introducing these names, we gain the power to define a general description operator:

$$\iota x \,.\, \psi(x) \quad \equiv \quad \bigcup\{y \,.\, x \in \{\emptyset\}, \psi(y)\}$$

Observe the peculiar usage of Replacement. The formula $\psi(y)$ is single-valued provided there exists a unique $a$ satisfying $\psi(a)$, and $\iota x \,.\, \psi(x)$ equals $a$. (If there exists no such $a$ then $\iota x \,.\, \psi(x)$ equals $\emptyset$, although this fact matters little.)

Because it demands uniqueness, $\iota x \,.\, \psi(x)$ is much weaker than Hilbert's description $\epsilon x \,.\, \psi(x)$, which embodies a strong version of the Axiom of Choice. Unique descriptions are still useful, as we shall see; their properties are summed up by two derived rules:

$$\frac{\psi(a) \qquad \begin{matrix}[\psi(x)]_x \\ \vdots \\ x = a\end{matrix}}{(\iota x \,.\, \psi(x)) = a}\ (\iota{=}) \qquad\qquad \frac{\exists! x \,.\, \psi(x)}{\psi(\iota x \,.\, \psi(x))}\ (\iota I)$$

## 7.3 Ordered pairs

The definition $\langle a, b\rangle \equiv \{\{a\}, \{a, b\}\}$ is perhaps the most famous (or notorious) feature of set theory. Isabelle defines

$$\langle a, b\rangle \equiv \{\{a, a\}, \{a, b\}\},$$

which is equivalent but consists entirely of doubletons. This simplifies the proof — which we shall examine later — of the key property

$$\langle a, b\rangle = \langle c, d\rangle \leftrightarrow a = c \wedge b = d.$$

The next step is to define the projections, `fst` and `snd`. Descriptions are extremely useful here. We could put

$$\begin{aligned} \texttt{fst}(p) &\equiv \iota x \,.\, \exists y \,.\, p = \langle x, y\rangle \\ \texttt{snd}(p) &\equiv \iota y \,.\, \exists x \,.\, p = \langle x, y\rangle \end{aligned}$$

To show $\texttt{fst}(\langle a, b\rangle) = a$ by the rule $(\iota{=})$, we must exhibit a unique $x$ such that $\exists y \,.\, \langle a, b\rangle = \langle x, y\rangle$ holds. Clearly $x = a$ (with $y = b$) by uniqueness of pairing. The treatment of `snd` is similar. Descriptions are suitable for defining many other kinds of destructors, such as case analysis operators for disjoint unions, natural numbers and lists. Isabelle's classical reasoner can prove the resulting equations.

Isabelle's ZF actually defines fst and snd indirectly. Following Martin-Löf's Constructive Type Theory [33], it defines the variable-binding projection $\texttt{split}(f,p)$, and proves the equation

$$\texttt{split}(f, \langle a,b\rangle) = f(a,b).$$

Frequently split is more convenient than the usual projections, which we define using meta-level $\lambda$-abstraction:

$$\begin{aligned} \texttt{fst}(p) &\equiv \texttt{split}(\lambda x\, y\,.\, x,\, p) \\ \texttt{snd}(p) &\equiv \texttt{split}(\lambda x\, y\,.\, y,\, p) \end{aligned}$$

Like other destructors, split is defined using a description:

$$\texttt{split}(f,p) \equiv \iota z\,.\,\exists x\, y\,.\, p = \langle x,y\rangle \wedge z = f(x,y).$$

## 7.4 Cartesian products

The set $A \times B$ consists of all pairs $\langle a,b\rangle$ such that $a \in A$ and $b \in B$. Many authors [21, 51] define the Cartesian product in a cumbersome manner. If $a \in A$ and $b \in B$ then $\{\{a\},\{a,b\}\} \in \wp(\wp(A \cup B))$, so they define $A \times B$ using Separation:

$$A \times B \equiv \{z \in \wp(\wp(A \cup B))\,.\,\exists_{x\in A}\,.\,\exists_{y\in B}\,.\; z = \langle x,y\rangle\}$$

There is a historical and pedagogical case for this definition, which postpones the introduction of Replacement. But Replacement is built into our notation, so we might as well take advantage of it:

$$A \times B \equiv \bigcup_{x\in A}\bigcup_{y\in B}\{\langle x,y\rangle\}$$

This definition is self-evident, independent of the underlying representation of pairs, and easy to reason about.

Again, Isabelle actually defines $A \times B$ indirectly, following Martin-Löf's Type Theory. The disjoint union of a family of sets, $\sum_{x\in A} .B(x)$, is a useful generalization of $A \times B$. To generalize the definition above, we merely replace $B$ by $B(x)$:

$$\sum_{x\in A} B(x) \equiv \bigcup_{x\in A}\bigcup_{y\in B(x)}\{\langle x,y\rangle\}$$

Natural deduction rules neatly summarize its properties:

$$\frac{a \in A \quad b \in B(a)}{\langle a,b\rangle \in (\sum_{x\in A} .B(x))}\ (\textstyle\sum I)$$

$$\frac{c \in (\sum_{x\in A} .B(x)) \qquad \begin{matrix}[x \in A \quad y \in B(x) \quad c = \langle x,y\rangle]_{x,y} \\ \vdots \\ \theta\end{matrix}}{\theta}\ (\textstyle\sum E)$$

By $(\sum E)$, if $\langle a,b\rangle \in (\sum_{x\in A} .B(x))$ then $a \in A$ and $b \in B(a)$.

Now $A \times B$ is nothing but an abbreviation for $\sum_{x\in A} .B(x)$ when $B$ has no dependence upon $x$. Isabelle's parser and pretty printer handle these conventions, using an ML function to search for occurrences of the bound variable.

## 7.5 Relations and functions

A **binary relation** is a set of ordered pairs. Isabelle's set theory defines the basic operations upon relations. These operations have the usual properties and require little discussion. Observe the usage of Replacement:

$$\begin{aligned}
\texttt{converse}(r) &\equiv \{z \,.\, w \in r,\, \exists x\, y \,.\, w = \langle x, y\rangle \wedge z = \langle y, x\rangle\} \\
\texttt{domain}(r) &\equiv \{x \,.\, w \in r,\, \exists y \,.\, w = \langle x, y\rangle\} \\
\texttt{range}(r) &\equiv \texttt{domain}(\texttt{converse}(r)) \\
\texttt{field}(r) &\equiv \texttt{domain}(r) \cup \texttt{range}(r)
\end{aligned}$$

**Image** and **inverse image** are infix operators:

$$\begin{aligned}
r \,``\, A &\equiv \{y \in \texttt{range}(r) \,.\, \exists_{x\in A} \,.\, \langle x, y\rangle \in r\} \\
r -`` A &\equiv \texttt{converse}(r)``A
\end{aligned}$$

**Functions** are represented by their graphs, which are single-valued binary relations. The set of all functions from $A$ to $B$ is written $A \to B$. Just as we generalized $A \times B$ to $\sum_{x\in A} .B(x)$, we generalize $A \to B$ to $\prod_{x\in A} .B(x)$, the product of a family of sets. This concept predates Martin-Löf's Type Theory; it has a long history. We define

$$\prod_{x\in A} B(x) \equiv \{f \in \wp(\Sigma_{x\in A} \,.\, B(x)) \,.\, \forall_{x\in A} \,.\, \exists! y \,.\, \langle x, y\rangle \in f\}.$$

Here $A \to B$ abbreviates $\prod_{x\in A} .B(x)$ when $B$ involves no dependence upon $x$. In particular, we have

$$(f \in A \to B) \leftrightarrow f \subseteq A \times B \wedge (\forall_{x\in A} \,.\, \exists! y \,.\, \langle x, y\rangle \in f).$$

We further define application and $\lambda$-abstraction. An explicit application operator is necessary; $f`a$ operates on the sets $f$ and $a$. Observe how easily a description expresses the application operator:

$$\begin{aligned}
f`a &\equiv \iota y \,.\, \langle a, y\rangle \in f \\
\lambda_{x\in A} \,.\, b(x) &\equiv \{\langle x, b(x)\rangle \,.\, x \in A\}
\end{aligned}$$

Regarding functions as binary relations requires proving many tiresome lemmas, such as

$$\frac{f \in (\prod_{x\in A} .B(x))}{\langle a, b\rangle \in f \leftrightarrow a \in A \wedge f`a = b}$$

It takes much reasoning of this sort to derive high-level rules for functions, in the style of the $\lambda$-calculus.

$$\frac{\begin{array}{c}[x \in A]_x \\ \vdots \\ b(x) \in B(x)\end{array}}{(\lambda_{x\in A} \,.\, b(x)) \in (\prod_{x\in A} .B(x))}\ (\lambda\Pi I) \qquad \frac{f \in (\prod_{x\in A} .B(x)) \quad a \in A}{f`a \in B(a)}\ (\lambda\Pi E)$$

$$\frac{a \in A}{(\lambda_{x\in A} \,.\, b(x))`a = b(a)}\ (\beta) \qquad \frac{f \in (\prod_{x\in A} .B(x))}{(\lambda_{x\in A} \,.\, f`x) = f}\ (\eta)$$

Injections, surjections and bijections are subsets of the total function space $A \to B$. Isabelle's set theory also defines composition of relations and functions:

$$\begin{aligned}
\texttt{inj}(A,B) &\equiv \{f \in A \to B \,.\, \forall_{w\in A} \,.\, \forall_{x\in A} \,.\ f`w = f`x \to w = x\} \\
\texttt{surj}(A,B) &\equiv \{f \in A \to B \,.\, \forall_{y\in B} \,.\, \exists_{x\in A} \,.\ f`x = y\} \\
\texttt{bij}(A,B) &\equiv \texttt{inj}(A,B) \cap \texttt{surj}(A,B) \\
r \circ s &\equiv \{w \in \texttt{domain}(s) \times \texttt{range}(r) \,. \\
&\qquad\qquad \exists x\, y\, z \,.\, w = \langle x,z\rangle \wedge \langle x,y\rangle \in s \wedge \langle y,z\rangle \in r\}
\end{aligned}$$

The numerous derived rules include

$$\frac{f \in \texttt{bij}(A,B)}{\texttt{converse}(f) \in \texttt{bij}(B,A)} \qquad \frac{f \in \texttt{inj}(A,B) \quad a \in A}{\texttt{converse}(f)`(f`a) = a}$$

$$\frac{s \subseteq A \times B \quad r \subseteq B \times C}{(r \circ s) \subseteq A \times C} \qquad \frac{g \in A \to B \quad f \in B \to C}{(f \circ g) \in A \to C}$$

$$(r \circ s) \circ t = r \circ (s \circ t)$$

Thus, relations and functions are closed under composition. A similar property is proved for injections, surjections and bijections.

# 8 Examples of set-theoretic reasoning

To give some idea of the level of reasoning possible in Isabelle, we shall examine three simple examples: ordered pairing, Cantor's Theorem, and the Composition of Homomorphisms challenge [9]. The sessions given below are based on polished proofs from Isabelle's set theory. I have simplified the commands to make the proofs slightly longer and easier to follow.

This section, which is intended for casual reading, describes the effect of each command in general terms. For details of the many Isabelle primitives that appear, please consult the documentation [38].

## 8.1 Injectivity of ordered pairing

Proving that $\langle a,b\rangle \equiv \{\{a,a\},\{a,b\}\}$ is a valid definition of ordered pairing is tiresome — see Halmos [21, page 23], for example. Here is a short machine proof using Isabelle's tools. We do not see all the details of a full proof (that happens internally) but we do see the key lemma. We now state this lemma, which concerns doubletons, to Isabelle:

```
goal ZF.thy "{a,b} = {c,d}  <->  (a=c & b=d) | (a=d & b=c)";
  Level 0
  {a,b} = {c,d} <-> a = c & b = d | a = d & b = c
   1. {a,b} = {c,d} <-> a = c & b = d | a = d & b = c
```

This is the initial state of a backward proof. It has one subgoal, which is the same as the main or ultimate goal. Our first inference will apply the derived rule

$$\frac{P \leftrightarrow Q \quad Q \leftrightarrow R}{P \leftrightarrow R}$$

to let us replace $\{a, b\} = \{c, d\}$ by any equivalent formula:

```
by (resolve_tac [iff_trans] 1);
  Level 1
  {a,b} = {c,d} <-> a = c & b = d | a = d & b = c
   1. {a,b} = {c,d} <-> ?Q
   2. ?Q <-> a = c & b = d | a = d & b = c
```

The one subgoal has become two, and the unknown intermediate formula appears as `?Q`. The first occurrence of $=$ in the main goal is one of the rare cases when the Axiom of Extensionality is directly useful. We replace $\{a, b\} = \{c, d\}$ by the inclusions $\{a, b\} \subseteq \{c, d\}$ and $\{c, d\} \subseteq \{a, b\}$, updating `?Q`.

```
by (resolve_tac [extension] 1);
  Level 2
  {a,b} = {c,d} <-> a = c & b = d | a = d & b = c
   1. {a,b} <= {c,d} & {c,d} <= {a,b} <-> a = c & b = d | a = d & b = c
```

Subgoal 1 has vanished; subgoal 2 has taken its place; `?Q` has become the conjunction of inclusions. The remaining subgoal requires a massive but essentially trivial case analysis. If $\{a, b\} \subseteq \{c, d\}$ then the rule $(\subseteq E)$ states that if $x \in \{a, b\}$ then $x \in \{c, d\}$; putting $x = a$ we obtain $a = c \vee a = d$, and so forth. (Halmos's proof makes a much smaller case analysis.) The classical tactic `fast_tac` proves the subgoal. It takes the collection of natural deduction rules proved so far, packaged as `upair_cs`.

```
by (fast_tac upair_cs 1);
  Level 3
  {a,b} = {c,d} <-> a = c & b = d | a = d & b = c
  No subgoals!
```

This automatic step takes about eight seconds.[4] Finally, we declare the resulting theorem as the ML identifier `doubleton_iff`:

```
val doubleton_iff = result();
```

Now we prove the main theorem, that ordered pairing is injective. While stating the goal, we make Isabelle expand the definition `Pair_def`:

```
goalw ZF.thy [Pair_def] "<a,b> = <c,d>  <->  a=c & b=d";
  Level 0
  <a,b> = <c,d> <-> a = c & b = d
   1. {{a,a},{a,b}} = {{c,c},{c,d}} <-> a = c & b = d
```

[4]All Isabelle timings are on a Sun SPARCstation ELC.

The expanded subgoal 1 is full of doubletons. We rewrite it using our lemma (FOL_ss is a collection of standard rewrite rules for first-order logic):

```
by (simp_tac (FOL_ss addsimps [doubleton_iff]) 1);
  Level 1
  <a,b> = <c,d> <-> a = c & b = d
   1. a = c & (b = d | c = d & b = d) |
      (a = c & c = d | a = d & d = c) & a = c & b = c <->
      a = c & b = d
```

The easiest way to prove the resulting subgoal involves further case analysis. This time, fast_tac requires only the rules of first-order logic, although supplying additional rules would do no harm.

```
by (fast_tac FOL_cs 1);
  Level 2
  <a,b> = <c,d> <-> a = c & b = d
  No subgoals!
```

Given the lemma, the total time to prove this theorem is about three seconds.

## 8.2 Cantor's Theorem

Cantor's Theorem is one of the few major results in mathematics that can be proved automatically. Its proof, although deep, is short. The prover TPS [2] can prove it in higher-order logic, where its statement is almost trivial. Some set theory systems can prove it too [4, 10].

```
goal ZF.thy "ALL f: A->Pow(A). EX S: Pow(A). ALL x:A. ~ f`x=S";
  Level 0
  ALL f:A -> Pow(A). EX S:Pow(A). ALL x:A. ~ f ` x = S
   1. ALL f:A -> Pow(A). EX S:Pow(A). ALL x:A. ~ f ` x = S
```

We begin by routine rule applications, using the introduction rules for the bounded quantifiers:

```
by (resolve_tac [ballI] 1);
  Level 1
  ALL f:A -> Pow(A). EX S:Pow(A). ALL x:A. ~ f ` x = S
   1. !!f. f : A -> Pow(A) ==> EX S:Pow(A). ALL x:A. ~ f ` x = S
```

Subgoal 1 requires showing $\exists_{S\in\wp(A)} . \forall_{x\in A} . f`x \neq S$ under the assumption $f \in A \to \wp(A)$, where $f$ is arbitrary.

```
by (resolve_tac [bexI] 1);
  Level 2
  ALL f:A -> Pow(A). EX S:Pow(A). ALL x:A. ~ f ` x = S
   1. !!f. f : A -> Pow(A) ==> ALL x:A. ~ f ` x = ?S1(f)
   2. !!f. f : A -> Pow(A) ==> ?S1(f) : Pow(A)
```

Under the same assumption, we now have two subgoals. The first, crucial goal involves the term `?S1(f)`, which is a placeholder for something that may depend upon $f$. Proving the subgoal instantiates this term with Cantor's diagonal set.

We can prove it automatically with `best_tac`, a classical reasoning tactic that employs best-first search. The search space is large and undirected. We must supply `best_tac` with a minimal collection of rules — though some readers might regard this as cheating. There are no heuristics for set theory; the diagonal set is found simply by brute force. Many of the subgoals consist largely of logical variables, for example $?a \in ?A$. The search resembles that reported by Bailin and Barker-Plummer [4].

```
val cantor_cs = FOL_cs
  addSIs [ballI, CollectI, PowI, subsetI] addIs [bexI]
  addSEs [CollectE, equalityCE];
```

Starting with `FOL_cs` — the rules for first-order logic — we add rules for the bounded quantifiers, powersets, the subset relation, Separation and extensional equality.

```
by (best_tac cantor_cs 1);
 Level 3
 ALL f:A -> Pow(A). EX S:Pow(A). ALL x:A. ~ f ` x = S
  1. !!f. f : A -> Pow(A) ==> {x: A . ~ x : f ` x} : Pow(A)
```

After five seconds, we have obtained the diagonal set, which is $\{x \in A . x \notin f`x\}$. The remaining subgoal is to show that the diagonal set belongs to $\wp(A)$. This is trivial; we may employ depth-first search (via `fast_tac`) and supply a large collection of rules (`ZF_cs`):

```
by (fast_tac ZF_cs 1);
  Level 4
  ALL f:A -> Pow(A). EX S:Pow(A). ALL x:A. ~ f ` x = S
  No subgoals!
```

Quaife [44, page 114] remarks that Otter could not construct the diagonal set; we have just seen Isabelle do so. Indeed, we could have proved Cantor's Theorem by a single call to `best_tac`. However, the classical reasoner is not designed to cope with such undirected searches. Equivalent forms of Cantor's Theorem cause the search to founder, even using the minimal collection of rules `cantor_cs`.

## 8.3 Composition of Homomorphisms

Boyer et al. [9] posed this as a challenge problem, and supplied a hand proof involving twenty-seven lemmas in BG set theory. Proving the theorem from the axioms alone might indeed be a challenge, but I found it easy in Isabelle's ZF set theory. The proof effort took about half an hour, much of which was spent keying in and correcting the conjecture. Most of the twenty-seven lemmas were already proved in Isabelle's set theory. Five concerned proving that certain classes were sets, which is necessary in BG but not in ZF. Other lemmas were perhaps proved on-the-fly by Isabelle's simplifier. My proof required no explicit lemmas.

The definition of homomorphism can be put into a more conventional notation (making the problem slightly harder!) by making $\mathtt{hom}(A, f, B, g)$ denote the set of all homomorphisms from $A$ to $B$:

$$\begin{aligned}\mathtt{hom}(A, f, B, g) \equiv \{ & H \in A \to B \,.\, (f \in A \times A \to A) \wedge (g \in B \times B \to B) \wedge \\ & (\forall_{x \in A} \,.\, \forall_{y \in A} \,.\, H`(f`\langle x, y\rangle) = g`\langle H`x, H`y\rangle)\}\end{aligned}$$

The contrast between the previous example and this one is clear. Cantor's Theorem is fundamental; its proof is short, but difficult to find. The proof that homomorphisms are closed under composition is straightforward, but long. The proof is mainly by rewriting, with some propositional reasoning to break up the conjunctions. We can set up `simp_tac` such that it calls `fast_tac` to prove its rewritten formulae, even when trying conditional rewrite rules; a single invocation of `simp_tac` proves the theorem in about forty seconds. But the proof is easier to follow if we perform it in several steps.

First we state the goal, binding the definition of homomorphism to the ML identifier `hom_def`:

```
val [hom_def] = goal Perm.thy
    "(!! A f B g. hom(A,f,B,g) ==                                  \
\       {H: A->B. f:A*A->A & g:B*B->B &                            \
\             (ALL x:A. ALL y:A. H`(f`<x,y>) = g`<H`x,H`y>)}) ==> \
\   J : hom(A,f,B,g) & K : hom(B,g,C,h) -->                        \
\   (K O J) : hom(A,f,C,h)";
  Level 0
  J : hom(A,f,B,g) & K : hom(B,g,C,h) --> K O J : hom(A,f,C,h)
   1. J : hom(A,f,B,g) & K : hom(B,g,C,h) --> K O J : hom(A,f,C,h)
```

Next, we expand `hom_def` in the subgoal:

```
by (rewtac hom_def);
  Level 1
  J : hom(A,f,B,g) & K : hom(B,g,C,h) --> K O J : hom(A,f,C,h)
   1. J :
      {H: A -> B .
       f : A * A -> A &
       g : B * B -> B &
       (ALL x:A. ALL y:A. H ` (f ` <x,y>) = g ` <H ` x,H ` y>)} &
      K :
      {H: B -> C .
       g : B * B -> B &
       h : C * C -> C &
       (ALL x:B. ALL y:B. H ` (g ` <x,y>) = h ` <H ` x,H ` y>)} -->
      K O J :
      {H: A -> C .
       f : A * A -> A &
       h : C * C -> C &
       (ALL x:A. ALL y:A. H ` (f ` <x,y>) = h ` <H ` x,H ` y>)}
```

Next we invoke a simple tactic from the classical reasoner, in order to break up conjunctions and remove the instances of Separation:

```
by (safe_tac ZF_cs);
  Level 2
  J : hom(A,f,B,g) & K : hom(B,g,C,h) --> K O J : hom(A,f,C,h)
   1. [| J : A -> B; K : B -> C; f : A * A -> A; g : B * B -> B;
         g : B * B -> B;
         ALL x:A. ALL y:A. J ` (f ` <x,y>) = g ` <J ` x,J ` y>;
         h : C * C -> C;
         ALL x:B. ALL y:B. K ` (g ` <x,y>) = h ` <K ` x,K ` y> |] ==>
      K O J : A -> C
   2. !!x y.
         [| J : A -> B; K : B -> C; f : A * A -> A; g : B * B -> B;
            g : B * B -> B;
            ALL x:A. ALL y:A. J ` (f ` <x,y>) = g ` <J ` x,J ` y>;
            h : C * C -> C;
            ALL x:B. ALL y:B. K ` (g ` <x,y>) = h ` <K ` x,K ` y>; x : A;
            y : A |] ==>
         (K O J) ` (f ` <x,y>) = h ` <(K O J) ` x,(K O J) ` y>
```

Next, we collect some rewrites to supply to the simplifier. The collection need not be minimal, so we begin with `ZF_ss` (a standard collection of rewrite rules) and add four relevant lemmas:

```
val hom_ss =
  ZF_ss addsimps [comp_func,comp_func_apply,SigmaI,apply_type];
```

Subgoal 1 is one of the lemmas, namely that functions are closed under composition.

Because simplification must employ the assumptions, in particular $J \in A \to B$ and $K \in B \to C$, the correct tactic here is `asm_simp_tac`:

```
by (asm_simp_tac hom_ss 1);
  Level 3
  J : hom(A,f,B,g) & K : hom(B,g,C,h) --> K O J : hom(A,f,C,h)
   1. !!x y.
         [| J : A -> B; K : B -> C; f : A * A -> A; g : B * B -> B;
            g : B * B -> B;
            ALL x:A. ALL y:A. J ` (f ` <x,y>) = g ` <J ` x,J ` y>;
            h : C * C -> C;
            ALL x:B. ALL y:B. K ` (g ` <x,y>) = h ` <K ` x,K ` y>; x : A;
            y : A |] ==>
         (K O J) ` (f ` <x,y>) = h ` <(K O J) ` x,(K O J) ` y>
```

Finally, we must show that $K \circ J$ maps applications of $f$ to applications of $h$. The simplifier applies the rewrite

$$\frac{g \in A \to B \quad f \in B \to C \quad a \in A}{(f \circ g)`a = f`(g`a)}$$

and uses the quantified assumptions about $J$ and $K$ as further rewrites. These rewrites are all conditional. The simplifier verifies the conditions using lemmas and the assumptions; this is essentially type checking.

```
by (asm_simp_tac hom_ss 1);
  Level 4
  J : hom(A,f,B,g) & K : hom(B,g,C,h) --> K O J : hom(A,f,C,h)
  No subgoals!
```

The total execution time for this proof is about seven seconds. Plaisted and Potter's system can prove this theorem in BG [42]; they also use rewriting.

# 9 Ramsey's Theorem in ZF

Ramsey's Theorem is a profound generalization of the pigeon-hole principle. A special case of it, the finite exponent 2 version, has become something of a benchmark for theorem provers. Basin and Kaufmann [5] compare proofs of this result using the Boyer/Moore Theorem Prover (called NQTHM) and Nuprl. The theorem is an informative example because its proof is both deep and long, involving graphs, sets and natural numbers. It covers a broad spectrum of reasoning issues. It is no toy example, but a major theorem with serious applications.

NQTHM and Nuprl differ in many respects. Isabelle (with ZF) is much closer to Nuprl: both support interactive, goal-directed proof using tactics and tacticals; both employ full predicate logic and some form of set theory. But Nuprl implements Martin-Löf's Constructive Type Theory rather than classical set theory.

## 9.1 The natural numbers in Isabelle's set theory

In set theory, the natural number $n$ is the $n$-element set $\{0, \ldots, n-1\}$. The companion paper will describe the construction of the set of natural numbers, and the derivation of recursion and induction. Isabelle's set theory proves many facts in elementary arithmetic. Here is a summary of the things needed for Ramsey's Theorem. The addition and subtraction operators are $\oplus$ and $\ominus$ because $+$ and $-$ stand for disjoint union and set difference, respectively.

| | |
|---|---|
| `nat` | set of natural numbers |
| 0 | zero (identical to $\emptyset$, the empty set) |
| $m \oplus n$ | sum of the natural numbers $m$ and $n$ |
| $m \ominus n$ | difference of the natural numbers $m$ and $n$ |
| `succ`$(m)$ | the successor of $m$, namely $m \oplus 1$ |

## 9.2 The definitions in ZF

Rather than attempt to improve upon Basin and Kaufmann's description of Ramsey's Theorem, I briefly discuss the corresponding definitions. I have used these largely as abbreviations, rather than as abstract notions; in most of the Isabelle proofs, the definitions are expanded.

Basin and Kaufmann's version of the theorem requires the notion of undirected graph, whose edge set $E$ is a symmetric binary relation. Sets of *unordered* pairs, instead of symmetric relations, would be more in harmony with the general finite version of Ramsey's Theorem [45].

$$\mathtt{Symmetric}(E) \equiv \forall x\, y\,.\, \langle x, y\rangle \in E \rightarrow \langle y, x\rangle \in E$$

Let $V$ be a set of vertices and $E$ a symmetric edge relation. Then $C$ is a **clique** if $C \subseteq V$ and every pair of distinct nodes in $C$ is joined by an edge in $E$. Dually, $I$ is an **independent set** (or **anticlique**) if $I \subseteq V$ and no pair of distinct nodes in $I$ is joined by an edge in $E$.

$$\mathtt{Clique}(C, V, E) \equiv C \subseteq V \wedge (\forall_{x\in C}\,.\, \forall_{y\in C}\,.\, x \neq y \rightarrow \langle x, y\rangle \in E)$$
$$\mathtt{Indept}(I, V, E) \equiv I \subseteq V \wedge (\forall_{x\in I}\,.\, \forall_{y\in I}\,.\, x \neq y \rightarrow \langle x, y\rangle \notin E)$$

Most of my efforts went to proving results that properly belong to a theory of cardinality. The NQTHM and Nuprl proofs both represent finite sets by lists without repetitions. This representation has disadvantages: it does not handle infinite sets; many of the laws for union and intersection fail. But the cardinality of such a 'set' is simply its length and many facts can be proved by routine inductions. Isabelle's ZF (as of this writing) does not define the cardinality of a set. Fortunately, the Ramsey proof requires only the notion '$S$ has at least $n$ elements.' This is equivalent to 'there is an injection from $n$ to $S$' because the natural number $n$ has $n$ elements:

$$\mathtt{Atleast}(n, S) \equiv \exists f\,.\, f \in \mathtt{inj}(n, S)$$

Finally, we define a predicate for Ramsey numbers:

$$\begin{aligned}\texttt{Ramsey}(n,i,j) \quad\equiv\quad & \forall V\,E\,.\,\texttt{Symmetric}(E)\land\texttt{Atleast}(n,V)\to\\ & (\exists C\,.\,\texttt{Clique}(C,V,E)\land\texttt{Atleast}(i,C))\lor\\ & (\exists I\,.\,\texttt{Indept}(I,V,E)\land\texttt{Atleast}(j,I))\end{aligned}$$

Now Ramsey's Theorem is easily stated:

$$\frac{i\in\texttt{nat}\quad j\in\texttt{nat}}{\exists_{n\in\texttt{nat}}\,\texttt{Ramsey}(n,i,j)}$$

Originally I defined

$$\texttt{Graph}(V,E)\quad\equiv\quad(E\subseteq V\times V)\land\texttt{Symmetric}(E)$$

and put $\texttt{Graph}(V,E)$ instead of $\texttt{Symmetric}(E)$ in the definition of Ramsey, but this was a needless complication. Since $E$ is universally quantified, the assertion holds for all $E$, including those such that $E\subseteq V\times V$.

All the lemmas proved for Ramsey's Theorem — except five that have been moved to the general library — are discussed below. Many are proved automatically.

## 9.3 Cliques and independent sets

The classical reasoner (fast_tac) proves these four facts automatically, taking half a second in total.

$$\texttt{Clique}(\emptyset,V,E)\qquad\texttt{Indept}(\emptyset,V,E)$$

$$\frac{\texttt{Clique}(C,V',E)\quad V'\subseteq V}{\texttt{Clique}(C,V,E)}\qquad\frac{\texttt{Indept}(I,V',E)\quad V'\subseteq V}{\texttt{Indept}(I,V,E)}$$

## 9.4 Cardinality

The classical reasoner automatically proves (in under two seconds) that every set has at least zero elements:

$$\texttt{Atleast}(0,A)$$

A useful rule for induction steps is derived by six explicit rule applications:

$$\frac{\texttt{Atleast}(\texttt{succ}(m),A)}{\exists_{x\in A}\,.\,\texttt{Atleast}(m,A-\{x\})}$$

A property of subsets has a short proof, using a related fact about injections:

$$\frac{\texttt{Atleast}(n,A)\quad A\subseteq B}{\texttt{Atleast}(n,B)}$$

This rule for adding an element to a set (cons) is proved by five rule applications:

$$\frac{\texttt{Atleast}(m,B)\quad b\notin B}{\texttt{Atleast}(\texttt{succ}(m),\texttt{cons}(b,B))}$$

Using `fast_tac` and the previous two results quickly yields

$$\frac{\texttt{Atleast}(m, B - \{x\}) \quad x \in B}{\texttt{Atleast}(\texttt{succ}(m), B)}$$

The following theorem is the pigeon-hole principle for two pigeon-holes. Proving it took up most of the time I devoted to Ramsey's Theorem. The proof involves induction on $m$ and $n$ with several case analyses; it consists of a complex mixture of proof checking with the tools `fast_tac` and `asm_simp_tac`.

$$\frac{m \in \texttt{nat} \quad n \in \texttt{nat} \quad \texttt{Atleast}(m \oplus n, A \cup B)}{\texttt{Atleast}(m, A) \lor \texttt{Atleast}(n, B)}$$

## 9.5 Ramsey's Theorem: the inductive argument

Ramsey's Theorem requires a double induction. Using previous lemmas, `fast_tac` proves the two base cases automatically (taking under two seconds in total):

$$\texttt{Ramsey}(0, 0, j) \qquad \texttt{Ramsey}(0, i, 0)$$

Before we can tackle the induction step, we must prove three lemmas. The first is an instance of the pigeon-hole principle:

$$\frac{\texttt{Atleast}(m \oplus n, A) \quad m \in \texttt{nat} \quad n \in \texttt{nat}}{\texttt{Atleast}(m, \{x \in A \,.\, \neg P(x)\}) \lor \texttt{Atleast}(n, \{x \in A \,.\, P(x)\})}$$

The next two lemmas contain the key idea of Ramsey's Theorem. One gives a method of extending a certain independent set of size $j$ to one of size $\texttt{succ}(j)$; the other gives a similar method for cliques. Using the definitions of `Symmetric`, `Indept`, and `Clique`, the standard rules (`ZF_cs`), and a lemma above concerning `Atleast`, `fast_tac` proves both theorems automatically! Each proof takes over half a minute, accounting for most of the CPU time in the entire proof.

$$\frac{\texttt{Symmetric}(E) \quad \texttt{Indept}(I, \{z \in V - \{a\} \,.\, \langle a, z\rangle \notin E\}, E) \quad a \in V \quad \texttt{Atleast}(j, I)}{\texttt{Indept}(\texttt{cons}(a, I), V, E) \land \texttt{Atleast}(\texttt{succ}(j), \texttt{cons}(a, I))}$$

$$\frac{\texttt{Symmetric}(E) \quad \texttt{Clique}(C, \{z \in V - \{a\} \,.\, \langle a, z\rangle \in E\}, E) \quad a \in V \quad \texttt{Atleast}(j, C)}{\texttt{Clique}(\texttt{cons}(a, C), V, E) \land \texttt{Atleast}(\texttt{succ}(j), \texttt{cons}(a, C))}$$

The induction step of Ramsey's Theorem is tedious, even with all the lemmas. The proof involves a four-way case split, with many explicit rule applications as well as invocations of the classical reasoner:

$$\frac{\texttt{Ramsey}(m, \texttt{succ}(i), j) \quad \texttt{Ramsey}(n, i, \texttt{succ}(j)) \quad m \in \texttt{nat} \quad n \in \texttt{nat}}{\texttt{Ramsey}(\texttt{succ}(m \oplus n), \texttt{succ}(i), \texttt{succ}(j))}$$

Finally, we prove the Theorem itself. This involves performing the double induction, invoking lemmas for the base cases and induction step:

$$\frac{i \in \texttt{nat} \quad j \in \texttt{nat}}{\exists_{n \in \texttt{nat}} \texttt{Ramsey}(n, i, j)}$$

## 9.6 Discussion and comparison

The induction step and base cases constitute a PROLOG program for `Ramsey`$(n, i, j)$, which we may express in a functional style:

$$
\begin{aligned}
r(0, j) &= 0 \\
r(i, 0) &= 0 \\
r(i+1, j+1) &= r(i+1, j) + r(i, j+1) + 1
\end{aligned}
$$

Since $r(i, j)$ computes a number $n$ satisfying `Ramsey`$(n, i, j)$, it is called the **witnessing function** for Ramsey's Theorem. Basin and Kaufmann [5] obtain slightly different Ramsey numbers; the definitions reflect details of the proofs.

Nuprl expresses Ramsey's Theorem using quantifiers, as here. Since its logic is constructive, Nuprl can extract a witnessing function from the proof. NQTHM lacks quantifiers; it expresses Ramsey's Theorem in terms of a witnessing function, obtained from a hand proof. Both the Nuprl and NQTHM proofs involve additional witnessing functions, which map a graph of sufficient size to a clique or independent set. The Isabelle proof follows the same reasoning as Basin and Kaufman's proofs; it does not make essential use of classical logic. Because it is conducted in classical ZF set theory, there is no way of extracting such witnessing functions from the proof.

The table compares the NQTHM, Nuprl and Isabelle/ZF proofs:

| | NQTHM | Nuprl | Isabelle |
|---|---|---|---|
| # Tokens | 933 | 972 | 975 |
| # Definitions | 10 | 24 | 5 |
| # Lemmas | 26 | 25 | 17 |
| # Replay Time | 3.7 minutes (Sun 3/60) | 57 minutes (Symbolics 3670) | 2.2 minutes (SPARC ELC) |

The figures for Isabelle include all the definitions and lemmas given above, and their proofs; they exclude the five lemmas that were moved to the general library. The Isabelle proof has the fewest definitions and lemmas. But NQTHM has the shortest replay time, since a Sun SPARCstation ELC is three or four times faster than a Sun 3/60. Kaufmann took seven hours to find the NQTHM proof; Basin required twenty hours, plus a further sixty for library development [5]. I took about nine hours to develop the Isabelle proof, including all twenty-two original lemmas.

Tokens were counted, after removal of comments, by the Unix command

```
sed -e "s/[^A-Za-z0-9'_]/ /g" ramsey.ML | wc
```

This counts identifiers but not symbols such as `:` and `=`, and is therefore an underestimate. It counts seven tokens in `EX x:A. Atleast(m, A-{x})`. Basin and Kaufmann each used different methods for counting tokens in their proofs. Figure 4 gives a more pessimistic impression of the token density of Isabelle proofs. One theorem is proved automatically. Another, which is the main induction step, has the second longest proof of the entire effort. The third is Ramsey's Theorem itself, with its inductions on $i$ and $j$.

```
val prems = goalw Ramsey.thy [Symmetric_def,Clique_def]
    "[| Symmetric(E);  Clique(C, {z: V-{a}. <a,z>:E}, E);  a: V;    \
\       Atleast(j,C) |] ==>                                         \
\    Clique(cons(a,C), V, E) & Atleast(succ(j), cons(a,C))";
by (cut_facts_tac prems 1);
by (fast_tac (ZF_cs addSEs [Atleast_succI]) 1);
val Clique_succ = result();

val ram1::ram2::prems = goalw Ramsey.thy [Ramsey_def]
   "[| Ramsey(m,succ(i),j);  Ramsey(n,i,succ(j));  m:nat;  n:nat |] ==> \
\   Ramsey(succ(m#+n), succ(i), succ(j))";
by (safe_tac ZF_cs);
by (etac (Atleast_succD RS bexE) 1);
by (eres_inst_tac [("P1","%z.<x,z>:E")] (Atleast_partition RS disjE) 1);
by (REPEAT (resolve_tac prems 1));
(*case m*)
by (rtac (ram1 RS spec RS spec RS mp RS disjE) 1);
by (fast_tac ZF_cs 1);
by (fast_tac (ZF_cs addEs [Clique_superset]) 1); (*we have a Clique*)
by (safe_tac ZF_cs);
by (eresolve_tac (swapify [exI]) 1);
by (REPEAT (ares_tac [Indept_succ] 1));          (*make a bigger Indept*)
(*case n*)
by (rtac (ram2 RS spec RS spec RS mp RS disjE) 1);
by (fast_tac ZF_cs 1);
by (safe_tac ZF_cs);
by (rtac exI 1);
by (REPEAT (ares_tac [Clique_succ] 1));          (*make a bigger Clique*)
by (fast_tac (ZF_cs addEs [Indept_superset]) 1); (*we have an Indept*)
val Ramsey_step_lemma = result();

val prems = goal Ramsey.thy
    "i: nat ==> ALL j: nat. EX n:nat. Ramsey(n,i,j)";
by (nat_ind_tac "i" prems 1);
by (fast_tac (ZF_cs addSIs [nat_0_I,Ramsey00j]) 1);
by (rtac ballI 1);
by (nat_ind_tac "j" [] 1);
by (fast_tac (ZF_cs addSIs [nat_0_I,Ramsey0i0]) 1);
by (dres_inst_tac [("x","succ(j1)")] bspec 1);
by (REPEAT (eresolve_tac [nat_succ_I,bexE] 1));
by (rtac bexI 1);
by (rtac Ramsey_step_lemma 1);
by (REPEAT (ares_tac [nat_succ_I,add_type] 1));
val ramsey = result();
```

Figure 4: Part of the Isabelle proof of Ramsey's Theorem

Comparisons are difficult. There are discrepancies in the hardware, token counting methods, etc. Furthermore, each author of a proof was an expert with his system. We can hardly predict how the systems would compare if tested by novices. The proof requires familiarity with both the system and its library of theorems.

Given these reservations, what conclusions can we draw? Isabelle stands up against two extensively developed systems, despite its lack of arithmetic decision procedures and small size (about 9000 lines of Standard ML, excluding object-logic definitions and proofs). More importantly, a generic proof assistant stands up against two systems designed for specific logics. This demonstrates the viability of generic theorem proving.

# 10 A brief history of ZF in Isabelle

Isabelle has supported ZF set theory since its early days. My original version consisted of idiosyncratic axioms over the sequent calculus LK [36, page 382]. Isabelle's set theory was developed only up to ordered pairs.

Philippe Noël found both the axioms and the sequent calculus uncongenial. He adopted Suppes's axioms and natural deduction (then newly available). Noël went on to prove a large body of results: theorems about relations, functions, orderings, fixed points, recursion, and more [32]. His priority was to develop as much mathematics as possible, not to create short and elegant proofs. Many of his proofs comprised ten, fifty or even 100 tactic steps.

## 10.1 The formalization of set theory

Noël generally worked by expanding definitions, which sometimes resulted in large formulae. The alternative to expanding definitions is to derive additional lemmas or rules. Deriving natural deduction rules for set theory (§5) simplified many proofs.

The description $\iota x\,.\,\psi[x]$ is a name for the unique object $a$ satisfying $\psi[a]$, if such exists. Descriptions are seldom mentioned in the literature, yet they are much more convenient than direct calculations. We can define the first projection by

$$\texttt{fst}(p) \quad \equiv \quad \iota x\,.\,\exists y\,.\,p = \langle x, y\rangle$$

instead of Noël's

$$\texttt{fst}(p) \quad \equiv \quad \bigcup(\bigcap(p)).$$

The former definition is independent of the representation of ordered pairing; to show $\texttt{fst}(\langle a, b\rangle) = a$, we simply appeal to a previous theorem about the injectivity of $\langle a, b\rangle$. The latter definition requires proving $\bigcup(\bigcap(\{\{a, a\}, \{a, b\}\})) = a$. The second projection (snd) can be defined easily using $\iota$, but otherwise requires a complex expression.

Simplifying the use of Replacement (§6) afforded improvements to existing definitions. For instance, Noël defined the domain of a relation using Separation:

$$\texttt{domain}(r) \quad \equiv \quad \{x \in \bigcup(\bigcup(r))\,.\,\exists y\,.\,\langle x, y\rangle \in r\}$$

Taking advantage of the simpler Replacement, I adopted

$$\texttt{domain}(r) \equiv \{x \,.\, w \in r,\ \exists y \,.\, w = \langle x, y\rangle\}.$$

This is more concise, and is independent of the representation of ordered pair.[5]

## 10.2 Tool development

When Noël started his work, Isabelle's simplifier and classical reasoner were crude. Noël developed a tactic that could prove many of his simpler theorems by expanding definitions [32].

Much later, I modified Isabelle's classical reasoner to be generic. I extended Isabelle with ways of controlling the instantiation of unknowns, to help prevent subgoals like $t \in ?A$ from causing runaway instantiations (see §3.1). Tobias Nipkow installed his simplifier [31].

Although slower than specialized provers, Isabelle's tools are fast enough: they normally return in a few seconds. Because my proof style minimizes the expanding of definitions, defining new concepts does not make proofs slower.

Tools obviously improve user productivity; moreover, the resulting proofs are resilient. Proof checking causes brittleness: proofs 'break' (fail to replay) after the slightest change to a definition or axiom. Tools generally adapt to changes. For a striking example of resilience, recall the pigeon-hole principle:

$$\frac{m \in \texttt{nat} \quad n \in \texttt{nat} \quad \texttt{Atleast}(m \oplus n, A \cup B)}{\texttt{Atleast}(m, A) \vee \texttt{Atleast}(n, B)}$$

The lemma can be strengthened: replace $m \oplus n$ by $m \oplus n \ominus 1$, where $1 \equiv \texttt{succ}(0)$. When I did this, the previous proof (consisting of twenty-eight commands!) replayed perfectly. The nested inductions went precisely as before; the case analyses were identical. The $\cdots \ominus \texttt{succ}(0)$ caused no difficulties because all subgoals containing it were submitted to the simplifier, using a general collection of arithmetic rewrites. This was partly luck, but the new version of the pigeon-hole principle required only slight changes to the rest of Ramsey's Theorem.

# 11 Related work

Few people have applied set theory to verification. Indeed, there has been little work on proving theorems from the axioms of set theory. On the other hand, there has been much work on automating set-theoretic reasoning by specialized heuristics and algorithms. Below, I survey the main strands of research and discuss their connection with the present work. Noël's work has been discussed above.

---

[5]Unless $r$ is known to be a binary relation, $\{\texttt{fst}(w) \,.\, w \in r\}$ is not equivalent to $\texttt{domain}(r)$.

## 11.1 Verification using axiomatic set theory

Corella's work on mechanizing set theory is closely connected with Isabelle's [13]. His proof assistant Watson embeds the Zermelo-Fraenkel axioms in higher-order logic (HOL), not first-order logic as is normally done. HOL supports the definition of variable-binding operators, as well as reasoning with axiom and theorem schemes. Isabelle uses a weak fragment of HOL for these purposes, but Watson uses full HOL. The combination of ZF and HOL contains a lot of redundancy, since HOL alone contains virtually the whole of mathematics. Watson is not generic, but perhaps could be made so; it provides strong support for notational shorthands. Watson is intended for verification and Corella demonstrates it with a small hardware proof.

Saaltink is also applying axiomatic set theory to verification. The EVES theorem prover implements first-order logic, with the ZF axioms added [48]. EVES seems to lack support for higher-order or schematic reasoning, but has a powerful simplifier. The EVES reducer can expand definitions and reason in various domains. EVES has a built-in mechanism for appealing to instances of axiom schemes. The EVES library [46, 47] derives a computational logic resembling Isabelle's.

## 11.2 Other proofs from the axioms of set theory

Frank M. Brown [10] adopts Quine's variant of set theory, which (like BG) admits both sets and classes. Brown's prover implements first-order logic using sequent methods, and is largely domain-independent. Heuristics control the the expansion of quantifiers and definitions. This is perhaps the earliest prover to combine unification and bound variables, and to use the description $\iota x \,.\, \psi[x]$.

EXCHECK performs computer-aided instruction in set theory. It supports goal-directed proof checking and its resolution prover can be applied to subgoals. It can give advice and summarize proofs. EXCHECK contains considerable domain knowledge; McDonald and Suppes describe it as an expert system for set theory [29, page 336].

Plaisted and Potter [42] have generated a proof of the Composition of Homomorphisms challenge. They use a sequent-based theorem prover, coded in Prolog, and BG set theory. Their work concerns replacing predicates by their definitions without having to Skolemize quantifiers dynamically; instead, positive and negative forms of the predicate are Skolemized in a translation phase.[6] Their translations resemble Isabelle's natural deduction rules for set theory (§5) — their rule for reducing the subgoal $A \subseteq B$ is essentially $(\subseteq I)$, while their rule for reducing the subgoal $A \not\subseteq B$ is the contrapositive of $(\subseteq E)$.

Quaife's resolution proofs in BG set theory using Otter [44] are remarkable. In a classic paper, Bledsoe argued that hard theorems required domain-dependent knowledge, and that resolution was the wrong approach [6]. Quaife encodes domain-dependent knowledge in the form of previously proved theorems; UR-resolution ensures that they are only applied usefully. Quaife's success reflects two decades' advance in technology. It also reflects meticulous care: lemmas are sometimes stated in a technical form; weights are sometimes set. Equality reasoning using the rule $A \subseteq B \wedge B \subseteq A \rightarrow A = B$ is particularly tricky [44, page 106]. Bledsoe's remark that resolution proofs are not "human-like" still applies.

[6] Kaufmann uses a similar approach to extend NQTHM with predicates defined by quantification [23].

## 11.3 Heuristics and algorithms for set-theoretic reasoning

Pastre's heuristics for set theory are based on "observation and imitation of the mathematicians' methods" [34]. A graph represents the binary relations that hold within a formula. The axiomatic basis is apparently BG; some axioms are given explicitly, while others are built into the algorithms. Using its graphs, the program can perform intricate chains of reasoning.

Ontic uses a Lisp-like language for set theory expressions and proofs [16]. A fragment of this language is executable. Sets, classes, and recursive functions may be defined. The ZF axioms are built into a knowledge base of facts about set theory. Ontic appeals to definitions and theorems automatically; the user never refers to them by name. The aim is to let users develop mathematics, requiring no knowledge of Ontic's inference mechanisms.

Cantone [11] is one of several related papers on decision procedures for fragments of set theory. He treats propositional combinations of the formulae $x = y \cup z$, $x = y - z$, $x \in y$, $x = \{y\}$ and $x = \wp(y)$. Such decision procedures do not constitute a theorem prover for set theory, but they could be valuable components of one. In an appendix, Cantone illustrates how the decision procedures could lead to an automated proof of a theorem from topology.

Bledsoe has done much research on solving for set unknowns, finding maximal or minimal sets. An early paper [7] presents examples from analysis. The work does not apply to untyped set theory — instantiations have the form $\{x \,.\, \psi[x]\}$, which need not denote a set — but rather to typed set theory. In later work, Bledsoe and Feng [8] describe the method as a procedure for second-order logic. It could probably be generalized to higher-order logic.

Bailin and Barker-Plummer's $\mathcal{Z}$-match rule [4] finds instantiations of the form $\{x \in A \,.\, \psi[x]\}$ for set unknowns. Their treatment of unknowns has three key features: (1) it can instantiate unknowns incrementally; (2) several subgoals may share an unknown and contribute to its instantiation; (3) a dual of Skolemization handles quantifiers. This treatment of unknowns has been invented independently twice before, in $\lambda$Prolog [30] and Isabelle [35], two systems based on higher-order unification. $\lambda$Prolog is a logic programming language that can easily express theorem provers [15]. Isabelle tactics using the ZF rules for Separation, such as

```
eresolve_tac [CollectI, CollectD1, CollectD2]
```

have roughly the same effect as $\mathcal{Z}$-match.

Bailin's unification algorithm for set theory is based on higher-order unification [3]. Bailin has modified Huet's procedure [22] to handle the reduction that takes $t \in \{x \,.\, \psi[x]\}$ to $\psi[t]$. Although Bailin's procedure can instantiate set variables, Bailin and Barker-Plummer [4] remark that it cannot do so incrementally.

## 11.4 Comparison

Considering the proofs alone, Quaife [44] and Saaltink [46, 47] have the most in common with the present paper. Both have developed set theory as far as functions, starting from the axioms. Quaife uses Otter with the BG axioms; his methods look formidable to me, but are perhaps common in the resolution world. Saaltink uses EVES with the ZF axioms; his paper [47] consists mainly of commands with little commentary, but suggests a fairly high level of interaction.

Isabelle's strong point is its treatment of variable-binding operators. Recall the schematic definitions of the Replacement operators (§6); Part II of this paper [40] defines several new binding operators to handle recursion. Quaife forgoes the notation $\{x\,.\,\psi[x]\}$ because Otter cannot handle it easily; he expresses classes using BG primitives in an obscure manner [44, page 97]. Since sets of the form $\{x\,.\,\psi[x]\}$ arise frequently in specifications, this is an obstacle to applying Quaife's method for verification. EVES also has difficulties with variable binding; defining a function by $\lambda$-abstraction takes half a page of commands [46]. The Isabelle theory defines $\lambda$-abstraction and proves $\beta$-reduction, both schematically (§7.5).

Ironically, Isabelle may not be suitable for BG set theory. In BG, the notation $\{x\,.\,\psi[x]\}$ is not part of the formalization itself, but is a meta-notation justified by the meta-theorem of class existence. The proof is by structural induction on the formula $\psi[x]$.[7] Such meta-theorems cannot be proved in Isabelle except by explicitly formalizing the syntax of first-order logic within some other logic. We cannot even assume the class existence theorem as an additional axiom — it does not hold schematically for all formulae, only for those with no quantification over classes.

# 12 Conclusions

Isabelle's version of ZF set theory, with its definitions, derived rules and tools, has reached a mature state of development. Problems can be stated in a reasonably familiar notation and approached using high-level steps.

Isabelle's set theory records roughly 1000 theorems. This paper discusses the formal development, starting from the ZF axioms, of a calculus of sets, pairs, relations and functions. The companion paper [40] develops general principles for defining recursive data types, including the natural numbers — using, for the first time, the Axiom of Infinity! The resulting theory can express both specifications and computable functions.

**Acknowledgements.** Philippe Noël's version of set theory, modified by Martin Coen, was the starting point of the present theory. Tobias Nipkow made great contributions to Isabelle, including the simplifier. David Basin, Matt Kaufmann, Brian Monahan and Philippe Noël commented usefully on this work. Thanks are also due to Robert Boyer (the editor) and to the referees.

[7] See Gödel [17, pages 39–43]; Boyer et al. [9] sketch the proof.